\documentclass{article}
\usepackage{spconf,amsmath,amssymb,graphicx,hyperref,adjustbox,booktabs,multirow,etoolbox,caption}
\usepackage[inkscapelatex=false]{svg}
\makeatletter
\apptocmd{\thebibliography}{%
  \spaceskip=.75\fontdimen2\font plus \fontdimen3\font minus \fontdimen4\font
}{}{}
\makeatother

\title{TokenChain: A Discrete Speech Chain via Semantic Token Modeling}
\twoauthors
 {Mingxuan Wang}
	{School of Data Science\\
	The Chinese University of Hong Kong, Shenzhen\\
    mingxuanwang1@link.cuhk.edu.cn}
 {Satoshi Nakamura\sthanks{Corresponding Author}}
	{School of Data Science and School of AI\\
	The Chinese University of Hong Kong, Shenzhen\\
    snakamura@cuhk.edu.cn}
\begin{document}
% \ninept
%
\maketitle
\begin{abstract}
Machine Speech Chain, simulating the human perception-production loop, proves effective in jointly improving ASR and TTS. We propose TokenChain, a fully discrete speech chain coupling semantic-token ASR with a two-stage TTS: an autoregressive text-to-semantic model co-trained with ASR and a masked-generative semantic-to-acoustic model for synthesis only. End-to-end feedback across the text interface is enabled with straight-through argmax/Gumbel--Softmax and balanced with supervised ASR via dynamic weight averaging. Ablations examine optimal temperature schedules for in- and cross-domain transfer. Evaluation reveals TokenChain surpasses baseline accuracy 2–6 epochs earlier and yields 5–13\% lower equal-epoch error with stable T2S on LibriSpeech, and reduces relative ASR WER by 56\% and T2S WER by 31\% on TED-LIUM with minimal forgetting, showing that chain learning remains effective with token interfaces and models.
\end{abstract}
\begin{keywords}
speech chain, discrete speech tokens, straight-through estimation, ASR, TTS.
\end{keywords}
\section{Introduction}
\label{sec:intro}
Human speech is a bidirectional mapping between symbolic text and acoustic realizations; coupling perception and production improves learning. The machine speech chain operationalizes this by training automatic speech recognition (ASR) and text-to-speech (TTS) in a semi-supervised closed loop \cite{tjandra2017listening}. Prior work enabled backpropagation from TTS to ASR via straight-through (ST) estimators \cite{tjandra2019end} and achieved domain adaptation \cite{yue2021exploring}, underscoring this paradigm’s promise. These systems largely employed continuous intermediates (mel-spectrograms, waveforms), presenting an opportunity to harness the shift toward token-centric sequence modeling.
\vspace{3.5pt}

Aligned with this opportunity, speech processing increasingly uses discrete tokens that integrate naturally with language models \cite{cui2024recent}. Guo et al. \cite{guo2025recent} distinguish semantic tokens, quantized from self-supervised (SSL) representations, from acoustic tokens learned for waveform reconstruction via neural codecs and residual vector-quantized (RVQ) models. Acoustic codecs such as SoundStream \cite{zeghidour2021soundstream} attain high fidelity; yet purely reconstruction-driven objectives can underweight linguistic content. To mitigate this, semantic distillation applied to SpeechTokenizer \cite{zhang2023speechtokenizer} guides early RVQ layers toward SSL targets to establish a semantic–acoustic hierarchy.
%such as w2v-BERT \cite{chung2021w2v}, HuBERT \cite{hsu2021hubert}, and WavLM \cite{chen2022wavlm}%
%(VQ) \cite{gray1984vector}%

Within TTS, mel-based continuous intermediates with diffusion/flow/GAN vocoders remain strong baselines \cite{xie2024towards}. Yet discrete-token systems are competitive. Hierarchical models first predict high-level semantic tokens and then expand them into finer acoustic tokens, as exemplified by AudioLM \cite{borsos2023audiolm}. Masked generative transformers enable efficient mask-and-predict decoding for token sequences: SoundStorm \cite{borsos2023soundstorm} realizes semantic-to-acoustic generation in this manner, and MaskGCT \cite{wang2024maskgct} extends the approach to a two-stage text-to-semantic and semantic-to-acoustic framework.

For ASR, scaling in self-supervised learning has driven gains, and discretizing SSL features emerged as viable inputs. Early unit-based systems required language models to match log-mel baselines \cite{baevski2020effectiveness}; recent work quantizes SSL features into compact codebooks to achieve competitive WER with lower storage and I/O \cite{chang2023exploration}. These results suggest semantic token sequences are suitable carriers for recognition and, by symmetry, promising targets for text-conditioned generation.

We revisit the speech chain in a fully discrete setting that (i) employs SOTA-competitive discrete ASR/TTS components, (ii) preserves end-to-end feedback through discrete estimation, (iii) reinstantiates the speech chain principle for semi-supervised closed-loop training via mutual feedback.

\begin{figure}[t]
    \centering
    \includegraphics[width=\columnwidth]{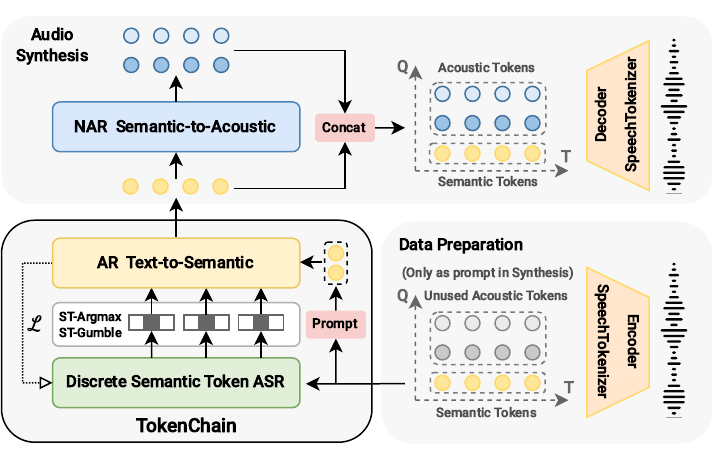}
    \caption{Illustration of proposed TokenChain framework. AR = autoregressive, NAR = non-autoregressive. The codec emits a $T\times Q$ token-index matrix ($Q$ = RVQ stages/codebooks).}
    % \vspace{-2mm}
    \label{fig:tokenchain}
\end{figure}

TokenChain (Fig. \ref{fig:tokenchain}) couples a discrete semantic-token ASR with a two-stage TTS: an autoregressive text-to-semantic (T2S) model co-trained with ASR, and a masked-generative semantic-to-acoustic (S2A) module for synthesis only. The ASR–T2S interface is textual, while feedback is entirely token-based: ST-argmax/Gumbel--Softmax enable backpropagation, and a semantic-token reconstruction loss is dynamically balanced with ASR cross-entropy. In chained training, TokenChain outperforms baselines, converging 2–6 epochs earlier with 5–13\% lower equal-epoch error, and under domain adaptation reduces WER by 56\% for ASR and 31\% for T2S with minimal forgetting, demonstrating that the speech-chain paradigm remains effective in the discrete-token era.

\section{Methods}
\label{sec:methods}

\subsection{Data Preparation}
\label{sec:data}
Transcripts are tokenized with byte-pair encoding (BPE) \cite{sennrich2015neural} into \(\mathbf{y}=(y_1,\dots,y_L)\) from a vocabulary of size \(C\). Speech is tokenized with SpeechTokenizer under semantic distillation: RVQ-1 is guided toward the layerwise mean of HuBERT \cite{hsu2021hubert} to concentrate linguistic content, while RVQ-2:8 capture residual acoustic detail. We denote RVQ-1 codes as semantic tokens \(\mathbf{s}=(s_1,\dots,s_T)\) and higher-layer stacks as acoustic tokens \(\mathbf{a}^{2:8}\) with \(\mathbf{a}^{j}=(a^j_1,\dots,a^j_T)\). Each utterance is encoded once; \((\mathbf{s},\mathbf{y})\) pair is used in TokenChain training, and a short acoustic prompt \(\mathbf{a}^p\) is retained only for audio synthesis.

\subsection{Discrete ASR (Semantic Tokens\texorpdfstring{$\rightarrow$}{→}Text)}
\label{sec:asr}
The ASR is an encoder-decoder with an optional CTC \cite{watanabe2017hybrid} branch. Given semantic token sequence $\mathbf{s}$, the decoder produces logits $\mathbf{h}_t^d\in\mathbb{R}^C$ and temperature-softmax posteriors
\begin{equation}
\label{eq:softmax-temp}
\mathbf{p}_y^t=\sigma_\tau(\mathbf{h}_t^d),\quad
\mathbf{p}_y^t[c]=\frac{\exp(\mathbf{h}_t^d[c]/\tau)}{\sum_{i=1}^{C}\exp(\mathbf{h}_t^d[i]/\tau)}.
\end{equation}
The attention CE and the CTC loss on encoder states $H$ are
% \begin{subequations}\label{eq:asr-losses}
% \begin{align}
% L_{\text{CE}}  &= -\frac{1}{L}\sum_{t=1}^{L}\log p_y^t[y_t], \\
% L_{\text{CTC}} &= -\log P_{\text{CTC}}(\mathbf{y}\mid H) \quad \text{\cite{graves2006connectionist}}
% \end{align}
% \end{subequations}
\begin{equation}
\label{eq:asr-ce}
L_{\text{CE}} = -\frac{1}{L}\sum_{t=1}^{L}\log \mathbf{p}_y^t[y_t].
\end{equation}
\begin{equation}
\label{eq:asr-ctc}
L_{\text{CTC}} = -\log P_{\text{CTC}}(\mathbf{y}\mid H).
\end{equation}
With CTC-attention hybrid formulation, the objective is
\begin{equation}
\label{eq:asr-hybrid}
L_{\text{ASR}}=(1-\eta)\,L_{\text{CE}}+\eta\,L_{\text{CTC}},\quad \eta\in[0,1].
\end{equation}

\subsection{Autoregressive T2S (Text\texorpdfstring{$\rightarrow$}{→}Semantic Tokens)}
\label{sec:t2s}
The T2S is a LLaMA-style \cite{touvron2023llama} causal LM that models
\begin{equation}
p_\theta(\mathbf{s}\mid \mathbf{P},\mathbf{s}^{p})=\prod_{t=1}^{T'} p_\theta\!\left(s_{t}\mid \mathbf{P},\mathbf{s}^{p}, s_{<t}\right),
\end{equation}
where $\mathbf{P}$ are text tokens and $\mathbf{s}^{p}$ is a semantic prefix prompt randomly sampled from the ground-truth $\mathbf{s}$ to convey speaker context, and $T'$ is the number of semantic target positions (excluding text and prompt). Training forms a prefix by concatenation $[\mathbf{P} \,\|\, \mathbf{s}^{p}]$ and computes cross-entropy only on the semantic targets (text and prompt positions are label-masked):
\begin{equation}
\label{eq:t2s-loss}
L_{\text{T2S}}=\frac{1}{\sum_t m_t}\sum_{t} m_t\,\mathrm{CE}\!\big(s_t,\hat{\mathbf{p}}_s^t\big),
\end{equation}
with $m_t\in\{0,1\}$ indicating semantic target positions and $\hat{\mathbf{p}}_s^t$ the T2S softmax over the semantic vocabulary. At inference, we condition on $[\mathbf{P}\,\|\,\mathbf{s}^{p}]$ and generate a semantic continuation. For chain training, T2S ingests embeddings corresponding to ASR outputs, enabling a differentiable link (below).

\subsection{Discrete Pass-Through and Chain Feedback}
\label{sec:feedback}
We close the loop by feeding ASR predictions to T2S and backpropagating a semantic reconstruction loss. Let $\hat{z}_t=\arg\max_c \mathbf{p}_y^t[c]$ and $\hat{\mathbf{y}}_t=\mathrm{onehot}(\hat{z}_t)$. We apply the straight-through (ST) estimator \cite{bengio2013estimating},
\begin{equation}
\label{eq:ste}
\frac{\partial \hat{\mathbf{y}}_t}{\partial \mathbf{p}_y^t}\approx \mathbf{I},\quad
\frac{\partial L_{\text{T2S}}}{\partial \theta_{\text{ASR}}}
\approx \sum_t \frac{\partial L_{\text{T2S}}}{\partial \hat{\mathbf{y}}_t}
\frac{\partial \mathbf{p}_y^t}{\partial \theta_{\text{ASR}}}.
\end{equation}
Additionally, for lower variance we use straight-through Gumbel--Softmax \cite{jang2016categorical}: draw $g_{t,c}\!\sim\!\mathrm{Gumbel}(0,1)$ and define
\begin{equation}
\label{eq:gumbel}
\tilde{\mathbf{p}}_y^t[c]=\frac{\exp\big((\mathbf{h}_t^d[c]+g_{t,c})/\tau\big)}{\sum_i \exp\big((\mathbf{h}_t^d[i]+g_{t,i})/\tau\big)}.
\end{equation}
In all experiments, the forward pass uses a hard one-hot $\hat{\mathbf{y}}_t$; in the backward pass, gradients propagate through ${\mathbf{p}}_y^t$ for ST-argmax and through $\tilde{\mathbf{p}}_y^t$ for ST-Gumbel.

\subsection{Final Objective with DWA Scheduling}
TokenChain's training objective is defined as:
\begin{equation}
\label{eq:final}
L_{\text{final}}=L_{\text{ASR}}+\alpha_e\,L_{\text{T2S}}.
\end{equation}
The chain weight $\alpha_e$ is scheduled by dynamic weight averaging (DWA) \cite{liu2019end} with warm-up. Let $L_k^{(e)}$ be the epoch-$e$ mean loss for $k\!\in\!\{\text{ASR},\text{T2S}\}$ and
$r_k^{(e)}=L_k^{(e)}/L_k^{(e-1)}$. With temperature $T$, the uncapped weight is
\begin{equation}
\label{eq:dwa}
\alpha_e^\star=\frac{\exp(r_{\text{T2S}}^{(e)}/T)}{\exp(r_{\text{ASR}}^{(e)}/T)+\exp(r_{\text{T2S}}^{(e)}/T)}.
\end{equation}
We use fixed warm-ups $\alpha_1=\alpha_{\mathrm{w0}}$, $\alpha_2=\alpha_{\mathrm{w1}}$; for $3\le e\le e_{\mathrm{ramp}}$, set a ramp $\alpha_e=\min(\alpha_e^\star,\alpha_{\max})$; afterwards $\alpha_e=\alpha_e^\star$.

\subsection{NAR S2A (Semantic\texorpdfstring{$\rightarrow$}{→}Acoustic Tokens)}
\label{sec:s2a}
For evaluation, we adopt a SoundStorm-style masked generative codec transformer conditioned on semantics to synthesize audio. Let \(\mathbf{a}^{2:8}\) be acoustic layers and \(\mathbf{a}^p\) a non-predicted short prompt. Per step, a target layer \(j\!\in\!\{2,\dots,8\}\) is sampled from a linear coarse-to-fine schedule; we apply a binary mask \(\mathbf{m}^j\) to obtain the partially observed \(\bar{\mathbf{a}}^{\,j}\) and predict
\begin{equation}
\label{eq:s2a}
p_{\phi}\!\left(\mathbf{a}^j \,\middle|\, \bar{\mathbf{a}}^{\,j},\; \mathbf{a}^{2:j-1},\; [\,\mathbf{a}^{p}\,\|\,\mathbf{s}\,]\right),
\end{equation}
using cross-entropy on masked tokens. Inputs embed the sum of lower acoustic layers $\mathbf{a}^{2:j-1}$ plus aligned semantic embeddings with the prompt $\mathbf{a}^{p}$ prepended. Inference generates layers sequentially ($j{=}2{\to}8$) via iterative parallel decoding.

\section{Experiments}
\label{sec:exp}

\subsection{Setup}
\label{sec:setup}
\textbf{Datasets.} We pretrain ASR/T2S on LibriSpeech-100 \cite{panayotov2015librispeech}, and run chain training on LibriSpeech-960 and TED-LIUM v2 \cite{rousseau2014enhancing}. Audio synthesis uses a trained-and-frozen S2A on Emilia \cite{he2024emilia}. We evaluate post-chain ASR/T2S on LibriSpeech dev/test-\{clean,other\} and TED-LIUM dev/test.
\vspace{1.5mm}

\noindent \textbf{Framework.} TokenChain framework is adopted from ESPnet \cite{watanabe2018espnet} for ASR/chain training and Amphion \cite{zhang2024amphion} for T2S/S2A training/evaluation to form a unified TokenChain pipeline.
\vspace{1.5mm}

\noindent \textbf{Metrics.}
For ASR, we evaluate CER and WER. For synthesized audio, we evaluate WER via Whisper-large-v3 \cite{radford2023robust}, speaker similarity (SIM-O) with WavLM–TDNN2 \cite{chen2022wavlm}, and speech quality using UTMOSv2 (Predicted MOS) \cite{baba2024utmosv2}.

\subsection{Model and Training Details}
\label{sec:impl}
\textbf{Discrete Semantic Token ASR.}
We use a joint CTC/attention encoder–decoder with an E-Branchformer encoder (12 blocks, 4 heads, cgMLP $k{=}31$, FFN dim 1024) and a 6-layer Transformer decoder (FFN dim 2048, dropout 0.1). The BPE text vocabulary is 5000. CTC weight $\eta{=}0.3$; no external LM. Decoding uses beam size 12 with CTC weight 0.3. For TokenChain, we resume from ASR/T2S checkpoint, keep optimizer/scheduler state, and override the ASR LR to $5{\times}10^{-4}$.
\vspace{-3mm}

\noindent \textbf{Autoregressive T2S.}
A LLaMA-style causal LM (§\ref{sec:t2s}) with $(\text{d}_{\text{model}}{=}1024,\ \text{intermediate\_size}{=}2048,\ \text{layers}{=}15)$; text vocabulary 5000, semantic token vocabulary 1027. Trained with AdamW (lr $2{\times}10^{-4}$), 32k warm-up inverse-sqrt schedule.
\vspace{2mm}

\noindent\textbf{Non-autoregressive S2A.}
A SoundStorm-style masked generative codec transformer (§\ref{sec:s2a}) with hidden size 1024, 16 layers, 16 heads, 7 quantizers (predicting RVQ-2:8; codebook 1024), CFG 0.15, linear mask-layer schedule. 32k warm-up inverse-sqrt schedule with AdamW (lr \(1{\times}10^{-4}\)).
\vspace{2mm}

\noindent \textbf{TokenChain Experiments.}
Starting from LibriSpeech-100 pretrained ASR/T2S, enable chain feedback (§\ref{sec:feedback}) on LibriSpeech-960 and TED-LIUM v2 for 20 epochs. As a comparison baseline, we train the same setup for 20 epochs with T2S feedback disabled. We compare ST-argmax, ST-Gumbel with temperature
$\tau\!\in\!\{\,\text{anneal }2.0{\to}0.1\text{ in 10 epochs},\allowbreak\,1.5,\,1.0,\,0.75\,\}$.
Early stopping halts training after 3 consecutive epochs without validation improvement.
The chain weight $\alpha_e$ follows DWA with warm-ups and capped ramp. In our runs $(\alpha_{\mathrm{w0}},\alpha_{\mathrm{w1}},\alpha_{\max},e_{\mathrm{ramp}},T)=(10^{-3},0.05,0.5,6,2)$

\subsection{Main Results on LibriSpeech}
\label{sec:results-ls}
Table~\ref{tab:ls960_asr} presents equal epoch budget (Epoch 12) CER/WER, when TokenChain variants have effectively converged and exceed baseline's final figures at Epoch 20. Across all splits, chain feedback outperforms the $L_{\text{ASR}}$-only baseline, adding to the baseline’s roughly halved error rate from the pre-chain checkpoint. The strongest variant, ST-Gumbel Anneal, yields further 10–13\% relative gains on clean sets and 5–9\% on other sets.  Fixed $\tau$ around 1.5 is competitive, whereas sharper $\tau{\le}1.0$ are less effective yet still surpass the baseline.

\begin{table}[h]
\centering
\small % >= 9pt for readability (scriptsize is usually too small)
\setlength{\tabcolsep}{2.8pt}
\renewcommand{\arraystretch}{0.95}
\captionsetup{skip=2pt,belowskip=0pt}
\caption{ASR Epoch-12 CER/WER (\%) on LibriSpeech-960.\\
d-\{c,o\}/t-\{c,o\} denotes dev-\{clean,other\}/test-\{clean,other\}.}
\vspace{1mm}
\label{tab:ls960_asr}
\begin{tabular}{lcccccccc}
\toprule
& \multicolumn{4}{c}{CER ↓} & \multicolumn{4}{c}{WER ↓} \\
\cmidrule(lr){2-5}\cmidrule(lr){6-9}
Model & d-c & d-o & t-c & t-o & d-c & d-o & t-c & t-o \\
\midrule
\emph{Pre-chain} (Epoch 0) & 4.0 & 10.5 & 4.0 & 10.9 & 10.4 & 23.1 & 10.6 & 23.9 \\
\midrule
Baseline ($L_{\text{ASR}}$ only) & 1.6 & 5.6 & 1.7 & 6.0 & 4.8 & 13.0 & 5.0 & 13.8 \\
\addlinespace[3pt]
ST-Argmax            & 1.5 & 5.3 & 1.5 & 5.7 & 4.4 & 12.5 & 4.5 & 13.2 \\
\textbf{ST-Gumbel Anneal} & \textbf{1.4} & \textbf{5.3} & \textbf{1.4} & \textbf{5.5} & \textbf{4.2} & \textbf{12.1} & \textbf{4.4} & \textbf{12.8} \\
ST-Gumbel 1.5         & 1.4 & 5.3 & 1.5 & 5.5 & 4.2 & 12.2 & 4.5 & 12.8 \\
ST-Gumbel 1.0         & 1.5 & 5.3 & 1.5 & 5.7 & 4.5 & 12.3 & 4.6 & 13.1 \\
ST-Gumbel 0.75        & 1.5 & 5.3 & 1.5 & 5.6 & 4.4 & 12.4 & 4.5 & 13.1 \\
\bottomrule
\end{tabular}
\end{table}

\noindent Table~\ref{tab:ls960_tts} evaluates T2S by calculating Whisper-WER, SIM-O, and predicted MOS on audio synthesized with a fixed S2A. Chain feedback with an apt $\tau$ can improve content robustness while preserving perceptual quality. ST-argmax achieves the lowest Whisper-WER (-11.6\% vs.\ baseline). SIM-O remains within $\pm$0.5 of baseline, and MOS is stable, peaking at $\tau{=}1.5$ (+0.06). Sharpening ST-Gumbel ($\tau{\le}1.0$) degrades WER with minimal change in SIM-O/MOS, indicating that sharp Gumbel interfaces (small $\tau$) impede text controllability of the T2S.
\vspace{-2mm}
\begin{table}[h]
\centering
\small % ~9pt in IEEEtran; larger than \scriptsize
\setlength{\tabcolsep}{4pt}
\renewcommand{\arraystretch}{0.95}
\captionsetup{skip=2pt,belowskip=0pt}
\caption{LibriSpeech TTS: Accuracy (WER \%), Speaker similarity (SIM-O), and Naturalness (Predicted MOS).}
\vspace{1mm}
\label{tab:ls960_tts}
\begin{tabular}{lccc}
\toprule
Model & WER ↓ & SIM-O ↑ & Pred. MOS ↑ \\
\midrule
\emph{Pre-chain} / Baseline  & 11.78 & 64.58 & 3.38 \\
\midrule
ST-Argmax     & \textbf{10.41} & 64.39 & 3.39 \\
ST-Gumbel Anneal & 12.73 & 64.94 & 3.41 \\
ST-Gumbel 1.5  & 11.37 & 64.72 & \textbf{3.44} \\
ST-Gumbel 1.0  & 13.40 & \textbf{65.05} & 3.39 \\
ST-Gumbel 0.75 & 15.52 & 64.40 & 3.41 \\
\bottomrule
\end{tabular}
\end{table}
\vspace{-2mm}

\newpage

\begin{figure}[t]
    \centering
    \includegraphics[width=\columnwidth]{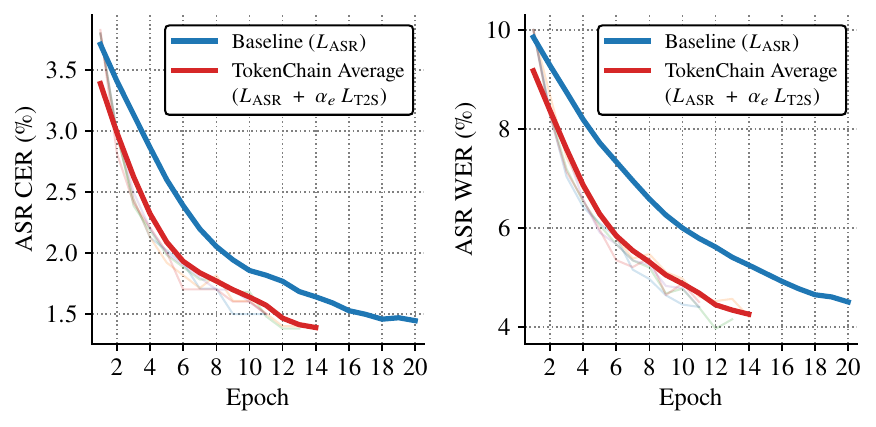}
    \captionsetup{skip=1pt,belowskip=0pt}
    \caption{Plot of LibriSpeech learning curves: CER (left), WER (right). TokenChain (red) vs.\ baseline (blue): surpasses the baseline 2–6 epochs earlier and achieves lower final error.}
    \label{fig:baseline-avg}
    \vspace{-2.5mm}
\end{figure}

\noindent \textbf{Convergence Efficiency.} Figure~\ref{fig:baseline-avg} shows that TokenChain’s learning curves are consistently below the baseline for CER and WER, surpassing baseline accuracy 2–6 epochs earlier. At Epoch 12, TokenChain already exceeds baseline’s final (Epoch 20) correct rates by 0.2 on clean sets and 0.4 on other sets. Hence, comparable or better accuracy is achieved with approximately 40\% fewer epochs (and compute), yielding substantial efficiency gains and underscoring chain feedback's role as a capable optimization aid and regularizer.

\subsection{Domain Adaptation on TED-LIUM}

\label{sec:results-ted}
Table~\ref{tab:ted_asr} reports final-epoch CER/WER, as the dominant TED-LIUM effect is improved generalization rather than earlier error-rate attainment. Although $L_{\text{ASR}}$-only baseline already cuts WER by 52\% from pre-chain, TokenChain provides further dev/test gains. The best setting, ST-Gumbel at $\tau{=}0.75$, attains 6.0/6.2 CER and 12.7/12.6 WER, corresponding to 55.3\% and 56.4\% total reductions, respectively. Argmax and annealed Gumbel are competitive but not optimal. The trend thus indicates that within ST-Gumbel, moderately sharper interfaces (smaller $\tau$) tend to facilitate cross-domain transfer.

\begin{table}[h]
\centering
\small % ~9pt in a 10pt base document
\setlength{\tabcolsep}{4pt}
\renewcommand{\arraystretch}{0.95}
\captionsetup{skip=2pt,belowskip=0pt}
\caption{ASR Final Epoch CER/WER (\%) on TED-LIUM.}
\vspace{1mm}
\label{tab:ted_asr}
\begin{tabular}{lcccc}
\toprule
& \multicolumn{2}{c}{CER ↓} & \multicolumn{2}{c}{WER ↓} \\
\cmidrule(lr){2-3}\cmidrule(lr){4-5}
Model & dev & test & dev & test \\
\midrule
\emph{Pre-chain} (Epoch 0) & 13.6 & 13.7 & 29.0 & 29.0 \\
\midrule
Baseline ($L_{\text{ASR}}$ only)   & 6.5 & 6.5 & 13.8 & 13.5 \\
\addlinespace[3pt]
ST-Argmax     & 6.1 & 6.4 & 12.8 & 13.0 \\
ST-Gumbel Anneal & 6.2 & 6.2 & 13.1 & 12.6 \\
ST-Gumbel 1.5  & 6.2 & 6.2 & 13.1 & 12.7 \\
ST-Gumbel 1.0  & 6.2 & 6.2 & 13.0 & 12.6 \\
\textbf{ST-Gumbel 0.75}  & \textbf{6.0} & \textbf{6.2} & \textbf{12.7} & \textbf{12.6}\\
\bottomrule
\end{tabular}
\end{table}

\begin{figure}[t]
    \centering
    \includegraphics[width=\columnwidth]{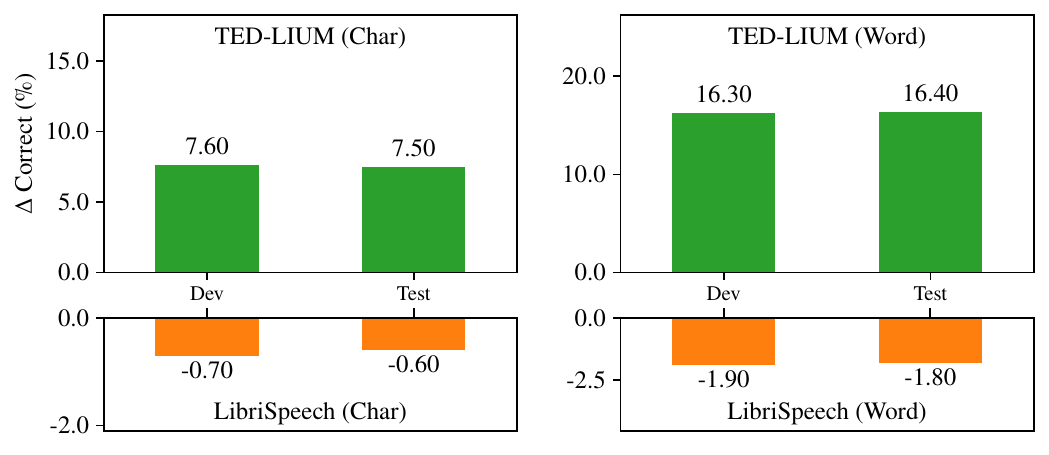}
    \captionsetup{skip=14pt,belowskip=0pt}
    \caption{Illustration of Domain Behavior: relative change in correct rate (higher is better). Top: major TED-LIUM gains; bottom: minor LibriSpeech performance degradations.}
    \label{fig:ted-lib}
    \vspace{-2.5mm}
\end{figure}

\noindent T2S results are summarized in Table~\ref{tab:ted_tts}: chaining yields consistent improvements over the pre-chain baseline across all criteria. Whisper-WER drops from 10.15 to 7.05–7.88, translating to $22$–$31\%$ relative gains, with best result occurring at $\tau{=}1.0$. Speaker similarity and MOS also improve by 4\% on average, with ST-argmax achieving the highest SIM-O (57.22) and predicted MOS (3.03). Thus, joint training transfers well to the target domain without sacrificing naturalness.

\begin{table}[h]
\centering
\small % ~9pt
\setlength{\tabcolsep}{4pt}
\renewcommand{\arraystretch}{0.95}
\captionsetup{skip=2pt,belowskip=0pt}
\caption{TED-LIUM TTS: Accuracy (WER \%), Speaker similarity (SIM-O), and Naturalness (Predicted MOS).}
\vspace{1mm}
\label{tab:ted_tts}
\begin{tabular}{lccc}
\toprule
Model & WER ↓ & SIM-O ↑ & Pred. MOS ↑ \\
\midrule
\emph{Pre-chain} / Baseline & 10.15 & 54.15 & 2.89 \\
\midrule
ST-Argmax     & 7.50  & \textbf{57.22} & \textbf{3.03} \\
ST-Gumbel Anneal & 7.85  & 56.56 & 3.00 \\
ST-Gumbel 1.5  & 7.88  & 56.81 & 2.98 \\
ST-Gumbel 1.0  & \textbf{7.05}  & 56.85 & 2.98 \\
ST-Gumbel 0.75 & 7.88  & 56.78 & 2.98 \\
\bottomrule
\end{tabular}
\end{table}

\noindent \textbf{Domain Behavior.} Figure~\ref{fig:ted-lib} quantifies the gain–loss asymmetry across domains: on TED-LIUM, correct rate gains are substantial: an increase of 7.5--7.6 on character and 16.3--16.4 on word, whereas on LibriSpeech, degradations are minimal: a decrease of 0.6--0.7 on character and 1.8--1.9 on word. This pattern indicates effective domain adaptation with limited forgetting of the source domain, showing that closed-loop chain feedback promotes domain-invariant semantic alignment.

% \subsection{Ablations on S2A}
% \label{sec:ablations-s2a}
% TBD

\section{Conclusions}
We introduced TokenChain, a discrete semantic-token closed-loop machine speech chain. Using ST-argmax/Gumbel--Softmax with dynamic weight averaging, it enables end-to-end feedback between a semantic-token ASR and an AR T2S while keeping a NAR S2A fixed for synthesis. Empirically, TokenChain improves recognition under equal epoch budget, converging 2–6 epochs earlier on LibriSpeech (5–13\% lower equal-epoch error); under TED-LIUM adaptation it achieves substantial ASR (56\%) and T2S WER (31\%) reductions with limited forgetting. Ablations show annealed ST-Gumbel ($\tau\!:\!\,2.0{\to}0.1$) is strongest in-domain, whereas a sharper interface ($\tau{\approx}0.75$) favors cross-domain transfer. Future work includes adaptive learning rate scheduling, joint S2A training, scaling to larger multilingual corpora, and human evaluations.

\vfill\pagebreak
\clearpage

\section{Acknowledgments}
This work was supported by the Guangdong Introducing Innovative and Entrepreneurial Teams Program.
\bibliographystyle{IEEEbib}
\bibliography{main}

\end{document}